# Quantum states with a space-like energy-momentum

by

Dan Solomon

Rauland-Borg
3450 W. Oakton
Skokie, IL

Please send all correspondence

to

Dan Solomon
1604 Brummel
Evanston, IL 60202

Email: dsolomon@northshore.net

July 11, 1999

**Abstract**

A common assumption in quantum field theory is that the energy-momentum 4-vector of any quantum state must be timelike. It will be proven that this is not the case for a Dirac-Maxwell field. In this case quantum states can be shown to exist whose energy-momentum is spacelike. This implies that there must exist quantum states with less energy than the vacuum state.

## 1. Introduction

In quantum field theory the eigenfunctions $|\varphi_n\rangle$ of the energy-momentum operator form a complete set of orthonormal basis states. Let $\hat{H}$ be the energy operator and $\hat{\vec{P}}$ be the momentum operator. The basis states $|\varphi_n\rangle$ satisfy,

$$\langle \varphi_m | \varphi_n \rangle = \delta_{mn} \tag{1}$$

and the energy and momentum eigenvalues are given by

$$\hat{H}|\varphi_n\rangle = E_n|\varphi_n\rangle; \quad \hat{\vec{P}}|\varphi_n\rangle = \vec{P}_n|\varphi_n\rangle \tag{2}$$

Any arbitrary state $|\Omega\rangle$ can be expressed as a series expansion in terms of the $|\varphi_n\rangle$, i.e.,

$$|\Omega\rangle = \sum_n g_n |\varphi_n\rangle \tag{3}$$

where $g_n$ are the expansion coefficients. (See standard works on quantum field theory such as Nishijima[1], Peskin and Schroeder [2], or Weinberg [3]). Throughout this discussion $|\Omega\rangle$ is assumed to be normalized, i.e., $\langle \Omega | \Omega \rangle = 1$.

For simple non-interacting systems equation (2) can be easily solved and the eigenvalues $E_n$ and $\vec{P}_n$ determined for each state $|\varphi_n\rangle$. For more complicated systems this is not the case. However, it is always assumed that the basis states $|\varphi_n\rangle$ and eigenvalues $E_n$ and $\vec{P}_n$ obey the following conditions. The first is that there is a lower bound to the energy eigenvalue. The eigenstate with the lowest energy eigenvalue is the vacuum state. Let $|\varphi_0\rangle$ be the vacuum state. The energy and momentum eigenvalues of the vacuum state are zero so that we can write

$$E_0 = 0 \text{ and } \vec{P}_0 = 0 \tag{4}$$

All other basis states are assumed to have a larger energy then the vacuum state so that we have

$$E_n > 0 \text{ for all } |\varphi_n\rangle \neq |\varphi_0\rangle \tag{5}$$

The second condition is that the energy-momentum 4-vector of each basis state must be timelike which means that,

$$E_n^2 - |\vec{P}_n|^2 \geq 0 \tag{6}$$

This actually follows from (5) and the special theory of relativity. Consider a frame $S'$ moving at a velocity $\vec{v}$ with respect to S where $|\vec{v}| < 1$. Here natural units are used so that the speed of light $c = 1$. The energy-momentum 4-vector of a basis state $|\varphi_n\rangle$ is $(E_n, \vec{P}_n)$. By a Lorentz transformation the energy in the frame $S'$ is given by

$$E'_n = \frac{E_n - \vec{v} \cdot \vec{P}_n}{\sqrt{1 - |\vec{v}|^2}} \tag{7}$$

(See p. 291 of M. Born [4]). If (6) is not true then it is possible to find a $|\vec{v}| < 1$ which makes $E'_n < 0$. This would contradict (5).

From the above discussion it can be shown that the energy-momentum 4-vector for any arbitrary state vector $|\Omega\rangle$ must be timelike, which means that,

$$\langle\Omega|\hat{H}|\Omega\rangle^2 - |\langle\Omega|\hat{\vec{P}}|\Omega\rangle|^2 \geq 0 \tag{8}$$

This can be shown as follows. Use (1), (2), and (3) to obtain

$$\langle\Omega|\hat{H}|\Omega\rangle = \sum_n |g_n|^2 E_n \tag{9}$$

and

$$\langle \Omega | \vec{P} | \Omega \rangle = \sum_n |g_n|^2 \vec{P}_n \tag{10}$$

The above two equations yield

$$\langle \Omega | \hat{H} | \Omega \rangle^2 = \sum_{nm} |g_n|^2 |g_m|^2 E_n E_m \tag{11}$$

and

$$\left| \langle \Omega | \vec{P} | \Omega \rangle \right|^2 = \sum_{nm} |g_n|^2 |g_m|^2 \vec{P}_n \cdot \vec{P}_m \tag{12}$$

Use the above to obtain

$$\langle \Omega | \hat{H} | \Omega \rangle^2 - \left| \langle \Omega | \vec{P} | \Omega \rangle \right|^2 = \sum_{nm} |g_n|^2 |g_m|^2 \left( E_n E_m - \vec{P}_n \cdot \vec{P}_m \right) \tag{13}$$

From (4), (5), and (6) it is obvious that $\left( E_n E_m - \vec{P}_n \cdot \vec{P}_m \right) \geq 0$ for all m and n. Therefore (8) is true.

Now a key element in the following discussion is that the relationships (5) and (6) are always assumed to be true. They are not proven or derived from more basic principles. It will be proven, for a quantum system consisting of a Dirac field interacting with an electromagnetic field, that the above condition is not true for all quantum states. It will be shown that, given an initial state $|\Omega\rangle$, that meets a certain requirement to be specified below, it is possible to find a state $|\Omega'\rangle$ for which the energy momentum 4-vector is spacelike, i.e.

$$\langle \Omega' | \hat{H} | \Omega' \rangle^2 - \left| \langle \Omega' | \hat{\vec{P}} | \Omega' \rangle \right|^2 < 0 \tag{14}$$

This, in turn, implies that there must exist at least one basis state $|\varphi_m\rangle$ for which

$$E_m^2 - \left| \vec{P}_m \right|^2 < 0 \tag{15}$$
5

A Lorentz transformation can then be done to yield a basis state with less energy than the vacuum state.

Throughout this discussion natural units are used so that $\hbar = c = 1$. Also the Schrodinger picture will be assumed. This means that the field operators are all time independent. The time dependence is reflected in the state vector which obeys the Schrodinger equation.

## 2. Energy-Momentum of Dirac-Maxwell system

It is shown in Appendix 1 that the energy operator $\hat{H}$ of the combined Dirac-Maxwell system in the coulomb gauge is given by

$$\hat{H} = \hat{H}_{0,D} + \hat{H}_I + \hat{H}_{CT} + \hat{H}_{0,EM} + \varepsilon_R \tag{16}$$

where,

$$\hat{H}_{0,D} = \int \left\{ -i\hat{\psi}^\dagger \vec{\alpha} \cdot \vec{\nabla}\hat{\psi} + m\hat{\psi}^\dagger \beta \hat{\psi} \right\} d\vec{x} \tag{17}$$

$$\hat{H}_I = -\int \hat{\vec{J}} \cdot \hat{\vec{A}} d\vec{x} \tag{18}$$

$$\hat{H}_{CT} = \frac{1}{2} \int d\vec{x} \int d\vec{x}' \frac{\hat{\rho}(\vec{x})\hat{\rho}(\vec{x}')}{4\pi|\vec{x} - \vec{x}'|} \tag{19}$$

$$\hat{H}_{0,EM} = \frac{1}{2} \int \left( \hat{\vec{E}}_\perp^2 + \hat{\vec{B}}^2 \right) d\vec{x} \tag{20}$$

In the above $\hat{\psi}$ is the Dirac field operator, $\vec{\alpha}$ and $\beta$ are the usual 4x4 matrices, $\hat{\vec{J}}$ is the current operator, $\hat{\rho}$ is the charge operator, $\hat{\vec{A}}$ is the operator for the vector potential, $\hat{\vec{E}}_\perp$ is the transverse electric field operator, $\hat{\vec{B}}$ is the magnetic field operator, and $\varepsilon_R$ is a renormalization constant so that the energy of the vacuum state is zero.

The momentum operator of the Dirac-Maxwell field is shown in Appendix 1 to be,

$$\hat{\vec{P}} = \hat{\vec{P}}_{O,D} + \hat{\vec{P}}_{0,EM} \tag{21}$$

where,

$$\hat{\vec{P}}_{O,D} = -i \int \psi^\dagger \vec{\nabla} \psi \, d\vec{x} \tag{22}$$

$$\hat{\vec{P}}_{0,EM} = \int \left( \hat{\vec{E}}_\perp \times \hat{\vec{B}} \right) d\vec{x} \tag{23}$$

From page 45 of P.W. Milonni [5] the electromagnetic field operators are given by

$$\hat{\vec{A}} = \sum_{\vec{k}\lambda} \left( \sqrt{\frac{1}{2|\vec{k}|V}} \right) \left( \hat{a}_{\vec{k}\lambda} e^{i\vec{k}\cdot\vec{x}} + \hat{a}^\dagger_{\vec{k}\lambda} e^{-i\vec{k}\cdot\vec{x}} \right) \vec{e}_{\vec{k}\lambda} \tag{24}$$

$$\hat{\vec{E}}_\perp = i \sum_{\vec{k}\lambda} \left( \sqrt{\frac{|\vec{k}|}{2V}} \right) \left( \hat{a}_{\vec{k}\lambda} e^{i\vec{k}\cdot\vec{x}} - \hat{a}^\dagger_{\vec{k}\lambda} e^{-i\vec{k}\cdot\vec{x}} \right) \vec{e}_{\vec{k}\lambda} \tag{25}$$

$$\hat{\vec{B}} = \vec{\nabla} \times \hat{\vec{A}} = i \sum_{\vec{k}\lambda} \left( \sqrt{\frac{1}{2|\vec{k}|V}} \right) \left( \hat{a}_{\vec{k}\lambda} e^{i\vec{k}\cdot\vec{x}} - \hat{a}^\dagger_{\vec{k}\lambda} e^{-i\vec{k}\cdot\vec{x}} \right) \vec{k} \times \vec{e}_{\vec{k}\lambda} \tag{26}$$

The $\hat{a}_{\vec{k}\lambda}$ and $\hat{a}^\dagger_{\vec{k}\lambda}$ are the photon annihilation and creation operators, respectively, for the mode with the wave vector $\vec{k}$ and polarization $\lambda$ where $\lambda = 1,2$. The $\vec{e}_{\vec{k}1}$ and $\vec{e}_{\vec{k}2}$ are unit polarization vectors which are perpendicular to each other and $\vec{k}$. V is the volume of the system which is allowed to approach infinity.

The photon annihilation and creation operators satisfy,

$$\left[ \hat{a}_{\vec{k}\lambda}, \hat{a}^\dagger_{\vec{k}'\lambda'} \right] = \delta_{\vec{k},\vec{k}'} \delta_{\lambda,\lambda'} \; ; \text{All other commutators=0} \tag{27}$$

They also commute with the Dirac field operators so that

$$\left[\hat{a}_{\vec{k}\lambda} \text{ or } \hat{a}^{\dagger}_{\vec{k}'\lambda}, \psi \text{ or } \psi^{\dagger}\right] = 0 \tag{28}$$

The following will be proved. Let $|\Omega\rangle$ be a state vector that satisfies (8) and the following condition

$$\int \vec{J}_e \cdot \vec{e}_{\vec{k}\lambda} \sin(\vec{k}\cdot\vec{x})d\vec{x} \neq 0 \tag{29}$$

for a given wave vector $\vec{k}$ and polarization $\lambda$, where $\vec{J}_e$ is the current expectation value defined by

$$\vec{J}_e = \langle\Omega|\hat{\vec{J}}|\Omega\rangle \tag{30}$$

Define the state

$$|\Omega'\rangle = e^{-if\hat{c}_{\vec{k}\lambda}}|\Omega\rangle \tag{31}$$

where f is a real number and $\hat{c}_{\vec{k}\lambda}$ is an operator defined by

$$\hat{c}_{\vec{k}\lambda} = \frac{1}{2}\left(\hat{a}^{\dagger}_{\vec{k}\lambda} + \hat{a}_{\vec{k}\lambda}\right) \tag{32}$$

We will prove that given (29) we can always find an f so that the energy-momentum 4-vector of the state $|\Omega'\rangle$ is spacelike.

The proof is very straightforward. Simply substitute (31) into the quantity $\langle\Omega'|\hat{H}|\Omega'\rangle^2 - \left|\langle\Omega'|\hat{\vec{P}}|\Omega'\rangle\right|^2$ and use the commutator relationships (27) and (28) along with the definition of the electromagnetic field operators (24), (25), and (26). At this point an expression is obtained which is dependent on the parameter f. It will be obvious from this expression that an f can always be found which makes the the energy-momentum of $|\Omega'\rangle$ spacelike.

Before proceeding consider the condition on the current given by (29). How do we know that a state $|\Omega\rangle$ exists where this is true? If our theory is a correct model of the real world then states must exist where (29) holds because (29) obviously can be true in classical physics which is often a very close approximation to the real world.

### 3. Mathematical details.

In this section the mathematical details required for the proof will be developed. Note that

$$\langle\Omega'| = \langle\Omega'|e^{if\hat{c}_{\vec{k}\lambda}^{\dagger}} \tag{33}$$

Also

$$\hat{c}_{\vec{k}\lambda}^{\dagger} = \hat{c}_{\vec{k}\lambda} \tag{34}$$

This yields

$$e^{if\hat{c}_{\vec{k}\lambda}^{\dagger}}e^{-if\hat{c}_{\vec{k}\lambda}} = e^{if\hat{c}_{\vec{k}\lambda}}e^{-if\hat{c}_{\vec{k}\lambda}} = 1 \tag{35}$$

Next we work out various commutator relationships which follow from (24) through (27). These are,

$$\left[\hat{\vec{A}}, \hat{a}_{\vec{k}\lambda}^{\dagger}\right] = \left(\sqrt{\frac{1}{2|\vec{k}|V}}\right)e^{+i\vec{k}\cdot\vec{x}}\vec{e}_{\vec{k}\lambda} \tag{36}$$

$$\left[\hat{\vec{A}}, \hat{a}_{\vec{k}\lambda}\right] = -\left(\sqrt{\frac{1}{2|\vec{k}|V}}\right)e^{-i\vec{k}\cdot\vec{x}}\vec{e}_{\vec{k}\lambda} \tag{37}$$

$$\left[\hat{\vec{E}}_{\perp}, \hat{a}_{\vec{k}\lambda}^{\dagger}\right] = i\left(\sqrt{\frac{|\vec{k}|}{2V}}\right)e^{+i\vec{k}\cdot\vec{x}}\vec{e}_{\vec{k}\lambda} \tag{38}$$

$$\left[\hat{\vec{E}}_\perp, \hat{a}_{\vec{k}\lambda}\right] = i\left(\sqrt{\frac{|\vec{k}|}{2V}}\right) e^{-i\vec{k}\cdot\vec{x}} \vec{e}_{\vec{k}\lambda} \tag{39}$$

$$\left[\hat{\vec{B}}, \hat{a}^\dagger_{\vec{k}\lambda}\right] = i\left(\sqrt{\frac{1}{2|\vec{k}|V}}\right) e^{+i\vec{k}\cdot\vec{x}} \vec{k}\times\vec{e}_{\vec{k}\lambda} \tag{40}$$

$$\left[\hat{\vec{B}}, \hat{a}_{\vec{k}\lambda}\right] = i\left(\sqrt{\frac{1}{2|\vec{k}|V}}\right) e^{-i\vec{k}\cdot\vec{x}} \vec{k}\times\vec{e}_{\vec{k}\lambda} \tag{41}$$

The above equations lead to the following expressions,

$$\left[\hat{\vec{A}}, \hat{c}_{\vec{k}\lambda}\right] = i\left(\sqrt{\frac{1}{2|\vec{k}|V}}\right) \vec{e}_{\vec{k}\lambda} \sin(\vec{k}\cdot\vec{x}) \tag{42}$$

$$\left[\hat{\vec{E}}_\perp, \hat{c}_{\vec{k}\lambda}\right] = i\left(\sqrt{\frac{|\vec{k}|}{2V}}\right) \vec{e}_{\vec{k}\lambda} \cos(\vec{k}\cdot\vec{x}) \tag{43}$$

$$\left[\hat{\vec{B}}, \hat{c}_{\vec{k}\lambda}\right] = i\left(\sqrt{\frac{1}{2|\vec{k}|V}}\right) \vec{k}\times\vec{e}_{\vec{k}\lambda} \cos(\vec{k}\cdot\vec{x}) \tag{44}$$

Next we want to evaluate the commutator $\left[\hat{O}, e^{-if\hat{c}_{\vec{k}\lambda}}\right]$ where $\hat{O} = \hat{\vec{A}}, \hat{\vec{E}}_\perp,$ or $\hat{\vec{B}}$.

Use the fact that $\left[\hat{O}, \hat{c}_{\vec{k}\lambda}\right]$ commutes with $\hat{c}_{\vec{k}\lambda}$ to obtain

$$\left[\hat{O}, \hat{c}^n_{\vec{k}\lambda}\right] = n\hat{c}^{n-1}_{\vec{k}\lambda}\left[\hat{O}, \hat{c}_{\vec{k}\lambda}\right] \tag{45}$$

Use this in the following Taylor's expansion

$$e^{-if\hat{c}_{\vec{k}\lambda}} = 1 + (-if)\hat{c}_{\vec{k}\lambda} + \frac{(-if)^2 \hat{c}^2_{\vec{k}\lambda}}{2!} + \ldots \frac{(-if)^n \hat{c}^n_{\vec{k}\lambda}}{n!} + \ldots \tag{46}$$

to obtain

$$\left[\hat{O}, e^{-if\hat{c}_{\vec{k}\lambda}}\right] = -ife^{-if\hat{c}_{\vec{k}\lambda}}\left[\hat{O}, \hat{c}_{\vec{k}\lambda}\right] \qquad (47)$$

Use this and (35) to yield

$$e^{if\hat{c}^{\dagger}_{\vec{k}\lambda}}\hat{O}e^{-if\hat{c}_{\vec{k}\lambda}} = e^{if\hat{c}^{\dagger}_{\vec{k}\lambda}}\left(e^{-if\hat{c}_{\vec{k}\lambda}}\hat{O} + \left[\hat{O}, e^{-if\hat{c}_{\vec{k}\lambda}}\right]\right) = \hat{O} - if\left[\hat{O}, \hat{c}_{\vec{k}\lambda}\right] \qquad (48)$$

Also, note that, from this and (35)

$$\begin{aligned} e^{if\hat{c}^{\dagger}_{\vec{k}\lambda}}\hat{O}_1\hat{O}_2 e^{-if\hat{c}_{\vec{k}\lambda}} &= \left(e^{if\hat{c}^{\dagger}_{\vec{k}\lambda}}\hat{O}_1 e^{-if\hat{c}_{\vec{k}\lambda}}\right)\left(e^{if\hat{c}^{\dagger}_{\vec{k}\lambda}}\hat{O}_2 e^{-if\hat{c}_{\vec{k}\lambda}}\right) \\ &= \left(\hat{O}_1 - if\left[\hat{O}_1, \hat{c}_{\vec{k}\lambda}\right]\right)\left(\hat{O}_2 - if\left[\hat{O}_2, \hat{c}_{\vec{k}\lambda}\right]\right) \end{aligned} \qquad (49)$$

Define the operator $\hat{D} = \hat{\vec{J}}, \hat{\rho}, \hat{H}_{0,D}$, or $\hat{H}_{CT}$. From (28) and (32) we obtain

$$\left[\hat{D}, e^{if\hat{c}_{\vec{k}\lambda}}\right] = 0 \qquad (50)$$

Define the following,

$$\vec{A}_{cl} = \left(\sqrt{\frac{1}{2|\vec{k}|V}}\right)\vec{e}_{\vec{k}\lambda}\sin(\vec{k}\cdot\vec{x}) \qquad (51)$$

$$\vec{E}_{cl} = \left(\sqrt{\frac{|\vec{k}|}{2V}}\right)\vec{e}_{\vec{k}\lambda}\cos(\vec{k}\cdot\vec{x}) \qquad (52)$$

$$\vec{B}_{cl} = \left(\sqrt{\frac{1}{2|\vec{k}|V}}\right)\vec{k}\times\vec{e}_{\vec{k}\lambda}\cos(\vec{k}\cdot\vec{x}) \qquad (53)$$

Use the above expressions along with (42), (43), and (44) in (48) to obtain

$$e^{if\hat{c}^{\dagger}_{\vec{k}\lambda}}\hat{\vec{A}}e^{-if\hat{c}_{\vec{k}\lambda}} = \hat{\vec{A}} + f\vec{A}_{cl} \qquad (54)$$

$$e^{if\hat{c}^{\dagger}_{\vec{k}\lambda}}\hat{\vec{E}}_{\perp}e^{-if\hat{c}_{\vec{k}\lambda}} = \hat{\vec{E}}_{\perp} + f\vec{E}_{cl} \qquad (55)$$

$$e^{if\hat{c}_{\vec{k}\lambda}^{\dagger}} \hat{\vec{B}} e^{-if\hat{c}_{\vec{k}\lambda}} = \hat{\vec{B}} + f\vec{B}_{cl} \tag{56}$$

## 4. Evaluate the energy-momentum

In the previous section we developed a number of expressions that will be useful in evaluating the momentum and energy of the state $|\Omega'\rangle$. First determine the momentum $\langle\Omega'|\hat{\vec{P}}|\Omega'\rangle$. Use (50), (35) and (33) to obtain

$$\langle\Omega'|\hat{\vec{P}}_{0,D}|\Omega'\rangle = \langle\Omega|\hat{\vec{P}}_{0,D}|\Omega\rangle \tag{57}$$

Next use (35) to obtain,

$$\langle\Omega'|\hat{\vec{P}}_{0,EM}|\Omega'\rangle = \langle\Omega|e^{-if\hat{c}_{\vec{k}\lambda}^{\dagger}}\left(\int \hat{\vec{E}}_{\perp} \times \hat{\vec{B}} d\vec{x}\right) e^{if\hat{c}_{\vec{k}\lambda}}|\Omega\rangle$$
$$= \langle\Omega|\left[\int\left(e^{-if\hat{c}_{\vec{k}\lambda}^{\dagger}} \hat{\vec{E}}_{\perp} e^{if\hat{c}_{\vec{k}\lambda}}\right) \times \left(e^{-if\hat{c}_{\vec{k}\lambda}^{\dagger}} \hat{\vec{B}} e^{if\hat{c}_{\vec{k}\lambda}}\right) d\vec{x}\right]|\Omega\rangle \tag{58}$$

Use (55) and (56) in the above to yield

$$\langle\Omega'|\hat{\vec{P}}_{0,EM}|\Omega'\rangle = \langle\Omega|\int\left(\left(\hat{\vec{E}}_{\perp} + f\vec{E}_{cl}\right) \times \left(\hat{\vec{B}} + f\vec{B}_{cl}\right)\right) d\vec{x}|\Omega\rangle \tag{59}$$

This yields,

$$\langle\Omega'|\hat{\vec{P}}_{0,EM}|\Omega'\rangle = \langle\Omega|\int\left(\hat{\vec{E}}_{\perp} \times \hat{\vec{B}}\right) d\vec{x}|\Omega\rangle + f\langle\Omega|\int\left(\hat{\vec{E}}_{\perp} \times \vec{B}_{cl}\right) d\vec{x}|\Omega\rangle$$
$$+ f\langle\Omega|\int\left(\vec{E}_{cl} \times \hat{\vec{B}}\right) d\vec{x}|\Omega\rangle + f^2\langle\Omega|\int\left(\vec{E}_{cl} \times \vec{B}_{cl}\right) d\vec{x}|\Omega\rangle \tag{60}$$

Define

$$\vec{E}_e = \langle\Omega|\hat{\vec{E}}_{\perp}|\Omega\rangle \text{ and } \vec{B}_e = \langle\Omega|\hat{\vec{B}}|\Omega\rangle \tag{61}$$

and use (23) to obtain

$$\langle\Omega'|\hat{\vec{P}}_{0,EM}|\Omega'\rangle = \langle\Omega|\hat{\vec{P}}_{0,EM}|\Omega\rangle + f\int\left\{\left(\vec{E}_e \times \vec{B}_{cl}\right) + \left(\vec{E}_{cl} \times \vec{B}_e\right)\right\} d\vec{x} + f^2\int\left(\vec{E}_{cl} \times \vec{B}_{cl}\right) d\vec{x} \tag{62}$$

From this, (57), and (21) we obtain

$$\langle\Omega'|\hat{\vec{P}}|\Omega'\rangle = \langle\Omega|\hat{\vec{P}}|\Omega\rangle + f\int\left\{\left(\vec{E}_e \times \vec{B}_{cl}\right) + \left(\vec{E}_{cl} \times \vec{B}_e\right)\right\}d\vec{x} + f^2\int\left(\vec{E}_{cl} \times \vec{B}_{cl}\right)d\vec{x} \qquad (63)$$

Next calculate the energy $\langle\Omega'|\hat{H}|\Omega'\rangle$. In analogy to the above discussion it is possible to write,

$$\langle\Omega'|\left(\hat{H}_{0,D} + \hat{H}_{CT} + \varepsilon_R\right)|\Omega'\rangle = \langle\Omega|\left(\hat{H}_{0,D} + \hat{H}_{CT} + \varepsilon_R\right)|\Omega\rangle \qquad (64)$$

and

$$\langle\Omega'|\hat{H}_I|\Omega'\rangle = -\langle\Omega|\int\hat{\vec{J}}\cdot\left(e^{if\hat{c}_{\vec{k}\lambda}^\dagger}\hat{\vec{A}}e^{-if\hat{c}_{\vec{k}\lambda}}\right)d\vec{x}|\Omega\rangle \qquad (65)$$

Use (54) in the above to obtain

$$\langle\Omega'|\hat{H}_I|\Omega'\rangle = -\langle\Omega|\int\hat{\vec{J}}\cdot\left(\hat{\vec{A}} + f\vec{A}_{cl}\right)d\vec{x}|\Omega\rangle \qquad (66)$$

Rearrange terms and use (18) to obtain

$$\langle\Omega'|\hat{H}_I|\Omega'\rangle = \langle\Omega|\hat{H}_I|\Omega\rangle - f\int\vec{J}_e\cdot\vec{A}_{cl}d\vec{x} \qquad (67)$$

Next evaluate,

$$\langle\Omega'|\hat{H}_{0,EM}|\Omega'\rangle = \langle\Omega|e^{if\hat{c}_{\vec{k}\lambda}^\dagger}\frac{1}{2}\int\left(\hat{\vec{E}}_\perp^2 + \hat{\vec{B}}^2\right)d\vec{x}\,e^{-if\hat{c}_{\vec{k}\lambda}}|\Omega\rangle \qquad (68)$$

Use (49), (55), and (56) to obtain

$$\langle\Omega'|\hat{H}_{0,EM}|\Omega'\rangle = \langle\Omega|\frac{1}{2}\int\left(\left|\hat{\vec{E}}_\perp + f\vec{E}_{cl}\right|^2 + \left|\hat{\vec{B}} + f\vec{E}_{cl}\right|^2\right)d\vec{x}\,|\Omega\rangle \qquad (69)$$

This yields

$$\langle\Omega'|\hat{H}_{0,EM}|\Omega'\rangle = \langle\Omega|\hat{H}_{0,EM}|\Omega\rangle + f\int\left\{\vec{E}_e\cdot\vec{E}_{cl} + \vec{B}_e\cdot\vec{B}_{cl}\right\}d\vec{x} + \frac{f^2}{2}\int\left(\vec{E}_{cl}^2 + \vec{B}_{cl}^2\right)d\vec{x} \qquad (70)$$

From (16) and (64) through (70) we obtain

$$\langle\Omega'|\hat{H}|\Omega'\rangle = \langle\Omega|\hat{H}|\Omega\rangle + f\left(\int\{\vec{E}_e\cdot\vec{E}_{cl} + \vec{B}_e\cdot\vec{B}_{cl}\}d\vec{x} - \int\vec{J}_e\cdot\vec{A}_{cl}d\vec{x}\right) + \frac{f^2}{2}\int\left(\vec{E}_{cl}^{\,2} + \vec{B}_{cl}^{\,2}\right)d\vec{x}$$

(71)

Now define

$$\eta = \int\{\vec{E}_e\cdot\vec{E}_{cl} + \vec{B}_e\cdot\vec{B}_{cl}\}d\vec{x} - \int\vec{J}_e\cdot\vec{A}_{cl}d\vec{x} \tag{72}$$

and

$$\xi_{cl} = \frac{1}{2}\int\left(\vec{E}_{cl}^{\,2} + \vec{B}_{cl}^{\,2}\right)d\vec{x} \tag{73}$$

Use these two expressions in (71) to obtain

$$\langle\Omega'|\hat{H}|\Omega'\rangle = \langle\Omega|\hat{H}|\Omega\rangle + f\eta + f^2\xi_{cl} \tag{74}$$

Similarly define

$$\vec{\chi} = \int\{(\vec{E}_e\times\vec{B}_{cl}) + (\vec{E}_{cl}\times\vec{B}_e)\}d\vec{x} \tag{75}$$

and

$$\vec{P}_{cl} = \int(\vec{E}_{cl}\times\vec{B}_{cl})d\vec{x} \tag{76}$$

Use this in (63) to obtain

$$\langle\Omega'|\hat{\vec{P}}|\Omega'\rangle = \langle\Omega|\hat{\vec{P}}|\Omega\rangle + f\vec{\chi} + f^2\vec{P}_{cl} \tag{77}$$

Take the square of (74) to obtain

$$\langle\Omega'|\hat{H}|\Omega'\rangle^2 = \langle\Omega|\hat{H}|\Omega\rangle^2 + f^2\eta^2 + f^4\xi_{cl}^2 + 2f\langle\Omega|\hat{H}|\Omega\rangle\eta \\ + 2f^2\langle\Omega|\hat{H}|\Omega\rangle\xi_{cl} + 2f^3\eta\xi_{cl} \tag{78}$$

Also square (77) to obtain

$$\left|\langle\Omega'|\hat{\vec{P}}|\Omega'\rangle\right|^2 = \left|\langle\Omega|\vec{P}|\Omega\rangle\right|^2 + f^2|\vec{\chi}|^2 + f^4|\vec{P}_{cl}|^2 + 2f\langle\Omega|\hat{\vec{P}}|\Omega\rangle\cdot\vec{\chi}$$
$$+ 2f^2\langle\Omega|\hat{\vec{P}}|\Omega\rangle\cdot\vec{P}_{cl} + 2f^3\vec{\chi}\cdot\vec{P}_{cl} \quad (79)$$

Therefore

$$\langle\Omega'|\hat{H}|\Omega'\rangle^2 - \left|\langle\Omega'|\hat{\vec{P}}|\Omega'\rangle\right|^2 = \left(\langle\Omega|\hat{H}|\Omega\rangle^2 - \left|\langle\Omega|\hat{\vec{P}}|\Omega\rangle\right|^2\right)$$
$$+ 2f\left(\langle\Omega|\hat{H}|\Omega\rangle\eta - \langle\Omega|\hat{\vec{P}}|\Omega\rangle\cdot\vec{\chi}\right)$$
$$+ f^2\left(\eta^2 + 2\langle\Omega|\hat{H}|\Omega\rangle\xi_{cl} - |\vec{\chi}|^2 - 2\langle\Omega|\hat{\vec{P}}|\Omega\rangle\cdot\vec{P}_{cl}\right)$$
$$+ 2f^3\left(\eta\xi_{cl} - \vec{\chi}\cdot\vec{P}_{cl}\right) + f^4\left(\xi_{cl}^2 - |\vec{P}_{cl}|^2\right) \quad (80)$$

Evaluate $\xi_{cl}$ using (52) and (53) along with the fact that $\int\left(\cos(\vec{k}\cdot\vec{x})\right)^2 d\vec{x} = V/2$ to obtain

$$\xi_{cl} = \frac{|\vec{k}|}{2V}\int\left(\cos(\vec{k}\cdot\vec{x})\right)^2 d\vec{x} = \frac{1}{4}|\vec{k}| \quad (81)$$

Also use the fact that $\vec{e}_{\vec{k}\lambda}\times(\vec{k}\times\vec{e}_{\vec{k}\lambda}) = \vec{k}$ to obtain

$$\vec{P}_{cl} = \frac{1}{2V}\left(\vec{e}_{\vec{k}\lambda}\times(\vec{k}\times\vec{e}_{\vec{k}\lambda})\right)\int\left(\cos(\vec{k}\cdot\vec{x})\right)^2 d\vec{x} = \frac{1}{4}\vec{k} \quad (82)$$

Therefore

$$\xi_{cl}^2 - |\vec{P}_{cl}|^2 = 0 \quad (83)$$

Now when this result is used in (80) it can be seen that the $f^4$ term is removed. This means that the highest order term in (80) is the $f^3$ term, as long as $\left(\eta\xi_{cl} - \vec{\chi}\cdot\vec{P}_{cl}\right)$ does not equal zero. If the sign of f is opposite that of $\left(\eta\xi_{cl} - \vec{\chi}\cdot\vec{P}_{cl}\right)$ then the quantity $2f^3\left(\eta\xi_{cl} - \vec{\chi}\cdot\vec{P}_{cl}\right)$ will be negative and if the magnitude of f is large enough, this term will dominate (80) making this expression negative. Therefore, it is possible to find an f which makes the energy-momentum 4-vector of the state $|\Omega'\rangle$ spacelike, as long as $\left(\eta\xi_{cl} - \vec{\chi}\cdot\vec{P}_{cl}\right)$ does not equal zero.

Let us examine the quantity $\left(\eta\xi_{cl} - \vec{\chi}\cdot\vec{P}_{cl}\right)$. It is shown in Appendix 2 that

$$\left(\eta\xi_{cl} - \vec{\chi}\cdot\vec{P}_{cl}\right) = -\xi_{cl}\int \vec{J}_e \cdot \vec{A}_{cl} d\vec{x} \tag{84}$$

As long as we pick the state $|\Omega\rangle$ so that (29) is true the above quantity will not be zero. As discussed previously we assume that this can always be done.

Given this result it is possible to show that there must exist at least one basis state $|\varphi_n\rangle$ for which $E_n^2 - \vec{P}_n^2 < 0$. The state $|\Omega'\rangle$ can be expanded in terms of the basis states per equation (3). Assume that we have chosen a value of f which makes (80) negative, then, from the discussion leading up to (13), we obtain

$$\langle\Omega'|\hat{H}|\Omega'\rangle^2 - \left|\langle\Omega'|\vec{P}|\Omega'\rangle\right|^2 = \sum_{nm}|g_n|^2|g_m|^2\left(E_n E_m - \vec{P}_n \cdot \vec{P}_m\right) < 0 \tag{85}$$

For this to be true there must be at least one term in the above sum which is negative. Assume that (5) is true so that all the $E_n$ are non-negative. Since $|g_n|^2|g_m|^2$ is non-negative then there must be at least one term where $\left(E_n E_m - \vec{P}_n \cdot \vec{P}_m\right) < 0$. If, for this case, n = m then we have that

$$E_n^2 - \left|\vec{P}_n\right|^2 < 0 \tag{86}$$

Suppose $n \neq m$. In this case

$$E_n E_m < \vec{P}_n \cdot \vec{P}_m \leq \left|\vec{P}_n\right|\left|\vec{P}_m\right| \tag{87}$$

For this to be true either $E_n < |\vec{P}_n|$ or $E_m < |\vec{P}_m|$ or both. In this case (86) holds for the index n or m (or both). In either case we can then do a Lorentz transformation to obtain a quantum state whose energy is less than that of the vacuum state.

## 5. Discussion

We have shown that there must exist quantum states whose energy-momentum 4-vector is spacelike. From this it follows that there must be basis states whose energy-momentum is also spacelike. A Lorentz transformation can, then, be done to produce a quantum state with negative energy with respect to the vacuum state.

Now the idea of a negative energy quantum state may seem like a disturbing concept to many physicist. To see why this should not be particularly disturbing let us consider what the requirements are for a vacuum state. One requirement is that the vacuum state must be a Lorentz invariant. The vacuum state must have the same properties in all inertial frames. Therefore the energy and momentum of the vacuum must be zero because if these quantities are both zero in one inertial frame they will be zero in all. Therefore equation (4) must be true.

Another requirement of the vacuum is that it must provide stability to positive energy particles. It is an observable fact that positive energy particles do not tend to decay into negative energy states. An assumption consistent with this is that these negative energy states do not exist and that the vacuum is the minimum energy state. This is the assumption that is normally made in quantum field theory. However, there is another assumption that is consistent with the observed facts and that is that these negative energy states do exist but that transitions into these states are difficult to achieve.

To illustrate this concept consider the non-relativistic Schrodinger equation in one dimensional space,

$$i\frac{\partial \psi(x,t)}{\partial t} = \left(-\frac{1}{2m}\frac{\partial^2}{\partial x^2} + V(x)\right)\psi(x,t) \qquad (88)$$

This equation describes the evolution of the wave function for a particle of mass m moving in a potential $V(x)$. Let

$$V(x) = \frac{1}{2}\omega^2 mx^2 - \lambda x^3 \qquad (89)$$

This describes the harmonic oscillator in the presence of a perturbing potential $-\lambda x^3$. Let us examine behavior of $V(x)$ for the case where $\lambda$ is small and positive. In the region where x is small $V(x)$ is dominated by the $x^2$ term. As we move out along the positive x-axis $V(x)$ will initially increase. However for a large enough x the $x^3$ will dominate and $V(x)$ will decrease and become negative.

The energy of a normalized wave function for this example is given by

$$E = \int \psi^\dagger \left(-\frac{1}{2m}\frac{\partial^2}{\partial x^2} + \frac{1}{2}\omega^2 mx^2 - \lambda x^3\right)\psi dx \qquad (90)$$

On examining this equation it can be seen that the first two terms in the integral will always be positive but that the integral over the $-\lambda x^3$ term can be negative depending on the value $\psi$. If $\psi$ is small for values of x away from the origin then the $x^3$ term will have a negligible effect on the energy. A physicist who "lives" in the region where x is small will observe that energy eigenstates and eigenvalues are close to that of a harmonic oscillator. He will also observe that there is ground state that is almost the same as the harmonic oscillator ground state and that all other states have greater energy then this ground state. However, if he explores his world for larger values of x he will find that the ground state is actually a local minimum. For wave functions whose magnitude is large for large positive values of x the $x^3$ term will dominate and the energy will be negative.

What this example illustrates is that it is possible to envision a stable system for which there exists negative energy states. What makes this system stable is that the negative energy levels are hard to get to from the ground state. There are two ways to get from the ground state to the negative energy states. One is by tunneling. However, for sufficiently small $\lambda$ the tunneling barrier will be very large so that the tunneling rate will be extremely small. The other way is to move over the potential barrier. This will require a large investment in energy to move the mass m up the potential energy slope in the direction of positive x. Therefore as long as our physicist confines his measurements to small values of x and small energy levels he will observe a stable world with a well defined ground state and positive energy levels above this ground state.

In conclusion, I believe that an analogous situation exists for the Dirac-Maxwell field. It has been shown that there must exist basis states with less energy than the vacuum state for a Dirac-Maxwell field. Transitions into these negative energy states must be difficult to achieve since they are not normally observed. However, for the theory to be mathematically complete, they must exist

### Appendix 1

The energy momentum tensor $\theta^{\mu\nu}$ of the Dirac-Maxwell field is discussed in Greiner & Reinhardt [6] (see equation 5 on page 151 of this reference). From this we can derive the total energy and momentum as

$$\hat{P}^0 = \hat{H}_{0,D} - \int \hat{\vec{J}} \cdot \hat{\vec{A}} d\vec{x} + \frac{1}{2} \int \left( \hat{\vec{E}}^2 + \hat{\vec{B}}^2 \right) d\vec{x} \tag{A1-1}$$

and

$$\hat{\vec{P}} = \hat{\vec{P}}_{0,D} - \int \hat{\rho} \hat{\vec{A}} d\vec{x} + \int \left( \hat{\vec{E}} \times \hat{\vec{B}} \right) d\vec{x} \tag{A1-2}$$

respectively (See chapters 5 and 6 of [6] and Appendix A of Sakurai [7]). The electromagnetic field will be quantized in the coulomb gauge. In this case

$$\vec{\nabla} \cdot \hat{\vec{A}} = 0 \tag{A1-3}$$

where

$$\hat{\vec{E}} = \left( \hat{\vec{E}}_\perp - \vec{\nabla} \hat{A}_0 \right) \text{ and } \hat{\vec{B}} = \vec{\nabla} \times \hat{\vec{A}} \tag{A1-4}$$

$\hat{\vec{E}}_\perp$ is the transverse component of the electric field and corresponds to the $-\partial \vec{A}/\partial t$ term of the classical unquantized electric field. $\hat{\vec{E}}_\perp$ satisfies.

$$\vec{\nabla} \cdot \hat{\vec{E}}_\perp = 0 \tag{A1-5}$$

In the coulomb gauge it can be shown that

$$\vec{\nabla}^2 \hat{A}_0 = -\hat{\rho} \tag{A1-6}$$

From Appendix A of Sakurai [7] it is shown that,

$$\frac{1}{2} \int \left( \hat{\vec{E}}^2 + \hat{\vec{B}}^2 \right) d\vec{x} = \frac{1}{2} \int \left( \hat{\vec{E}}_\perp^2 + \hat{\vec{B}}^2 \right) d\vec{x} + \frac{1}{2} \int d\vec{x} \int d\vec{x} \frac{\hat{\rho}(\vec{x}) \hat{\rho}(\vec{x}')}{4\pi |\vec{x} - \vec{x}'|} \tag{A1-7}$$

Use this in (A1-1). Then replace $\hat{P}^0$ by $\hat{H}$ to convert to the notation used in this discussion and add the renormalization constant $\varepsilon_R$ to obtain (16).

Next, use (A1-4) to obtain,

$$\int \left( \hat{\vec{E}} \times \hat{\vec{B}} \right) d\vec{x} = \int \left( \left( \hat{\vec{E}}_\perp - \vec{\nabla} \hat{A}_0 \right) \times \hat{\vec{B}} \right) d\vec{x} \tag{A1-8}$$

Next evaluate $\int \left( \vec{\nabla} \hat{A}_0 \times \hat{\vec{B}} \right) d\vec{x} = \int \left( \vec{\nabla} \hat{A}_0 \times \left( \vec{\nabla} \times \hat{\vec{A}} \right) \right) d\vec{x}$. From a vector identity,

$$\vec{\nabla} \left( \hat{\vec{A}} \cdot \vec{\nabla} \hat{A}_0 \right) = \left( \hat{\vec{A}} \cdot \vec{\nabla} \right) \vec{\nabla} \hat{A}_0 + \left( \vec{\nabla} \hat{A}_0 \cdot \vec{\nabla} \right) \hat{\vec{A}} + \vec{\nabla} \hat{A}_0 \times \left( \vec{\nabla} \times \hat{\vec{A}} \right) \tag{A1-9}$$

we obtain

$$\int \vec{\nabla} \hat{A}_0 \times \left( \vec{\nabla} \times \hat{\vec{A}} \right) d\vec{x} = -\int \left\{ \left( \vec{\nabla} \hat{A}_0 \cdot \vec{\nabla} \right) \hat{\vec{A}} + \left( \hat{\vec{A}} \cdot \vec{\nabla} \right) \vec{\nabla} \hat{A}_0 \right\} d\vec{x} \tag{A1-10}$$

where we have assumed reasonable boundary conditions so that

$$\int \vec{\nabla} \left( \hat{\vec{A}} \cdot \vec{\nabla} \hat{A}_0 \right) d\vec{x} = 0 \tag{A1-11}$$

Assume reasonable boundary conditions and integrate by parts to obtain

$$\int \vec{\nabla} \hat{A}_0 \times \left( \vec{\nabla} \times \hat{\vec{A}} \right) d\vec{x} = \int \left\{ \left( \vec{\nabla}^2 \hat{A}_0 \right) \hat{\vec{A}} + \left( \vec{\nabla} \cdot \hat{\vec{A}} \right) \vec{\nabla} \hat{A}_0 \right\} d\vec{x} \tag{A1-12}$$

Use (A1-3) and (A1-6) in the above expression to obtain

$$\int \vec{\nabla} \hat{A}_0 \times \left( \vec{\nabla} \times \hat{\vec{A}} \right) d\vec{x} = -\int \hat{\rho} \hat{\vec{A}} d\vec{x} \tag{A1-13}$$

Use the above expressions to obtain

$$\int \left( \hat{\vec{E}} \times \hat{\vec{B}} \right) d\vec{x} = \int \left( \hat{\vec{E}}_\perp \times \hat{\vec{B}} \right) d\vec{x} + \int \hat{\rho} \hat{\vec{A}} d\vec{x} \tag{A1-14}$$

Use this in (A1-2) to obtain (21) in the text.

## Appendix 2

Evaluate $\left( \eta \xi_{cl} - \vec{\chi} \cdot \vec{P}_{cl} \right)$. From (82), (81), (72), and (75) we obtain

$$\left(\eta\xi_{cl} - \vec{\chi}\cdot\vec{P}_{cl}\right) = \frac{1}{4}\left|\vec{k}\right|\left(\int\left\{\vec{E}_e\cdot\vec{E}_{cl} + \vec{B}_e\cdot\vec{B}_{cl}\right\}d\vec{x} - \int\vec{J}_e\cdot\vec{A}_{cl}d\vec{x}\right)$$
$$-\frac{1}{4}\vec{k}\cdot\int\left\{\left(\vec{E}_e\times\vec{B}_{cl}\right) + \left(\vec{E}_{cl}\times\vec{B}_e\right)\right\}d\vec{x} \qquad (A2\text{-}1)$$

Use (53) to obtain

$$\vec{k}\cdot\left(\vec{E}_e\times\vec{B}_{cl}\right) = \vec{k}\cdot\left(\vec{E}_e\times\left(\vec{k}\times\vec{e}_{\vec{k}\lambda}\right)\right)\left(\sqrt{\frac{1}{2\left|\vec{k}\right|V}}\right)\cos(\vec{k}\cdot\vec{x}) \qquad (A2\text{-}2)$$

Use the vector identity

$$\vec{E}_e\times\left(\vec{k}\times\vec{e}_{\vec{k}\lambda}\right) = \left(\vec{E}_e\cdot\vec{e}_{\vec{k}\lambda}\right)\vec{k} - \left(\vec{E}_e\cdot\vec{k}\right)\vec{e}_{\vec{k}\lambda} \qquad (A2\text{-}3)$$

in the above and recall that $\vec{k}\cdot\vec{e}_{\vec{k}\lambda} = 0$ to obtain

$$\vec{k}\cdot\left(\vec{E}_e\times\vec{B}_{cl}\right) = \left|\vec{k}\right|\left(\vec{E}_e\cdot\vec{e}_{\vec{k}\lambda}\right)\left(\sqrt{\frac{\left|\vec{k}\right|}{2V}}\right)\cos(\vec{k}\cdot\vec{x}) \qquad (A2\text{-}4)$$

Use (52) in the above to obtain

$$\vec{k}\cdot\left(\vec{E}_e\times\vec{B}_{cl}\right) = \left|\vec{k}\right|\left(\vec{E}_e\cdot\vec{E}_{cl}\right) \qquad (A2\text{-}5)$$

Next, use (52) to obtain

$$\vec{k}\cdot\left(\vec{E}_{cl}\times\vec{B}_e\right) = \vec{k}\cdot\left(\vec{e}_{\vec{k}\lambda}\times\vec{B}_e\right)\left(\sqrt{\frac{\left|\vec{k}\right|}{2V}}\right)\cos(\vec{k}\cdot\vec{x}) \qquad (A2\text{-}6)$$

Use the vector identity

$$\vec{k}\cdot\left(\vec{e}_{\vec{k}\lambda}\times\vec{B}_e\right) = \left(\vec{k}\times\vec{e}_{\vec{k}\lambda}\right)\cdot\vec{B}_e \qquad (A2\text{-}7)$$

and (53) to obtain

$$\vec{k}\cdot\left(\vec{E}_{cl}\times\vec{B}_e\right) = \left|\vec{k}\right|\vec{B}_{cl}\cdot\vec{B}_e \qquad (A2\text{-}8)$$

Use (A2-5) and (A2-8) in (A2-1) to obtain

$$\left(\eta\xi_{cl} - \vec{\chi}\cdot\vec{P}_{cl}\right) = \frac{1}{4}\left(-\int \vec{J}_e \cdot \vec{A}_{cl}d\vec{x}\right)\left|\vec{k}\right| \qquad (A2-9)$$

**References**

1. K. Nishijima, *Fields and Particles*, Benjamin/Cummings Publishing Co., Reading, Massachusetts (1974).

2. M. E. Peskin and D. V. Schroeder, *An Introduction to Quantum Field Theory*, Addison-Wesley Publishing Company, Reading Massachusetts (1995).

3. S. Weinberg, *The Quantum Theory of Fields Vol. 1*, Press Syndicate of the University of Cambridge, Cambridge, (1995).

4. M. Born, *Einstien's Theory of Relativity*, Dover Publications, Inc., New York (1965).

5. P. W. Milonni, *The Quantum Vacuum*, An Introduction to Quantum Electrodynamics, Acedamic Press, San Diego, CA (1994). These equations differ from those in Milonni by a factor of $\sqrt{1/4\pi}$. This is because Milonni includes a factor of $1/8\pi$ in the definition of the energy of the electromagnetic field. In this paper a factor of $1/2$ is used (see Eq. (20)).

6. W. Greiner and J. Reinhardt. *Quantum Electrodynamics*. Springer-Verlag, Berlin. (1992).

7. J. J. Sakurai, *Advanced Quantum Mechanics*, Addison-Wesley Publishing Co., Redwood, Calif (1967).